# Structural, electronic, and magnetic properties of ZnTe doped with transition metal Mn

A. BRACH, L. BAHMAD and S. BENYOUSSEF

Laboratory of Condensed Matter and Interdisciplinary Sciences (LaMCScI), Faculty of Sciences, Mohammed V University, Av. Ibn Batouta, B. P. 1014 Rabat, Morocco.

**ABSTRACT**

In this article, we examine the structure and the electronic, optical, and magnetic properties of ZnTe before and after doping with the transition metal Mn. The ab initio calculations of this compound were performed using the full potential linearized extended full potential planar waveform (FP-LAPW) in the context of density functional theory (DFT) implemented in the Wien2K code. The potential for exchange and correlation was addressed by the generalized gradient approximation (GGA) approximation. The electronic properties show that the ZnTe material exhibits semiconductor behavior before doping. As a result, it becomes semimetal after doping. The findings attained by Monte Carlo simulations display that the ZnMnTe material goes from an antiferromagnetic phase to the paramagnetic phase at the Neel temperature value $T_N = 159.31$ K.

## 1  Introduction

Condensed matter physics and scientific materials play an increasingly important role in technological applications which will only increase in many areas [1-6].

Before materials (solids) are used in industry, the quality of their structural, electronic, mechanical, and optical properties must be ensured [7]. The physical properties of a solid are closely related to the behavior of the electrons that constitute it [8-10]. The main goal of condensed matter theory is to solve the problem of the electronic structure of solids [11-12]. Electronic structure theory is useful both for understanding and interpreting experimental results, and for serving as a means of prediction [13-15]. For a fundamental understanding of electronic structure and materials, theorists have developed methods based on semi-empirical models [16]. Such models often include many parameters that can be adjusted to the experimental data.

---

[1] Corresponding author: bahmad@fsr.ac.ma

Other more rigorous and sophisticated computational methods called ab initio, based on the fundamental quantum theory that use only the atomic constants as input parameters for the solution of the Schrödinger equation [17-18].

These methods have become a basic tool for the study of structural, electronic, mechanical, and optical properties of molecules and materials. They are also a tool of choice for the study of certain effects that are difficult or impossible to determine experimentally, as well as, for the prediction of new materials because they have sometimes been able to replace very expensive or even unfeasible laboratory experiments.

The classification of materials depends on the desired application. We mainly rely on semiconductors for using optoelectronic or photovoltaic applications, which have aroused great interest in their experimental analysis and theoretical development. Monoatomic crystals whose leader and first representative are silicon place in the first rank of these semiconductors [19]. In fact, silicon is represented as an excellent candidate for various applications. It is synthesized with very high purity, and then extended later to binary compounds of the same lineage of the type GaAs with a structure known as Zinc Blende [20].

The race to integrate silicon-based microelectronics is driven by economic reasons, and it must now reach its limits [21]. In recent years, alternative solutions to silicon microelectronics have appeared. Various semiconductor compounds are attracting a lot of interest, starting with wide band gap semiconductors (GaN, AlN, SiC, ZnTe, ZnO, Diamond, etc.) which have been studied for several years in an absolute competitive technological context [22].

Zinc Telluride is a prototype II-VI semiconductor material with a direct gap of 2.26 eV [23]. It is usually a p-type semiconductor. It has a Blende structure like most compound semiconductor materials. In recent years, these semiconductors have attracted a lot of attention due to the direct band gap energy and the property of emitting light at room temperature. In addition, as the power of computers increases, it is easier to calculate the properties of materials, which are structural, electronic, and optical solids with great accuracy. These compounds are suitable for many technological applications, such as solid-state laser devices, photovoltaic devices, solar cells, remote control systems, thin films, transistors, detectors, imaging systems, etc [24-26]. The wavelength of the emitted radiation depends essentially on the Gap width of the material used. In the case of visible light-emitting diodes, the gap width must be between 1.8 eV and 2.6 eV [27]. Moreover, some other recent works using DFT method have been subject of equiatomic quaternary Heusler alloys ZnCdXMn (X=Pd, Ni or Pt) [28], Ab-initio calculations for the electronic and magnetic properties of Cr doped ZnTe [29], and Ab initio calculations of the magnetic properties of TM (Ti, V)-doped zinc-blende ZnO [30].

In this work, we use two methods; The full Potential Linearized Augmented Plane Wave (FPLAPW) method with GGA approximation and Monte Carlo simulations to study the structural, magnetic, and electronic properties of ZnTe and ZnMnTe. Additionally, Monte Carlo simulations are applied to derive the magnetic properties according to the Ising model.

## 2 Crystal structure and method

To calculate the structural, electronic, optical, and magnetic properties of ZnTe, we used the Wien2K package [31-32]. The Kohn-Sham equation system is solved using the full potential linearized extended plane wave (FPLAPW) [33].

ZnTe crystallized in the Zinc-Blende crystal structure (**Fig.1**). and belongs to the space group F43m(No.216) with lattice constants a = b = c = 6.101Å [34]. The compound ZnTe consists of two sub-arrays face-centered cubic sub-lattice shifted one concerning the other of a quarter of the large diagonal of the cube. The first sub-lattice is occupied by the cations of column II (Zn) while the other is occupied by the anions of column VI (Te) [27].

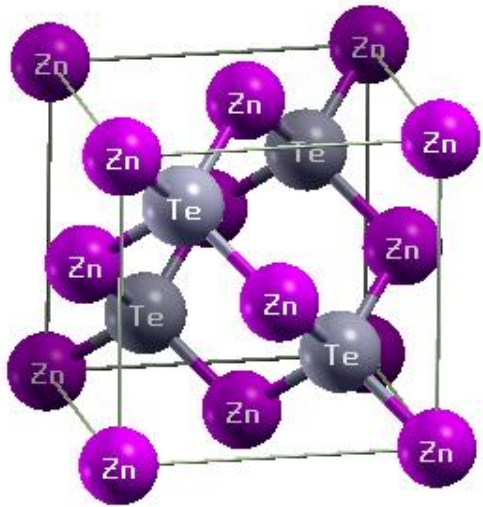

**Fig. 1** Zinc-Blende structure ZnTe

# 3  Results and discussion

## 3.1  Structural properties

To study the structural properties of the cubic compound ZnTe, we calculated the total energy as a function of unit cell volume around the equilibrium cell volume to determine the structural lattice constants using GGA approximation. The values of the experimental lattice parameters are used as the initial input while minimizing the total energy concerning the volume of the unit cell [27]. The results obtained from the total energies are presented as a function of the unit cell volume for the ZnTe compound (**Fig. 2**). The calculated equilibrium lattice constant of ZnTe is 6.205371 Å.

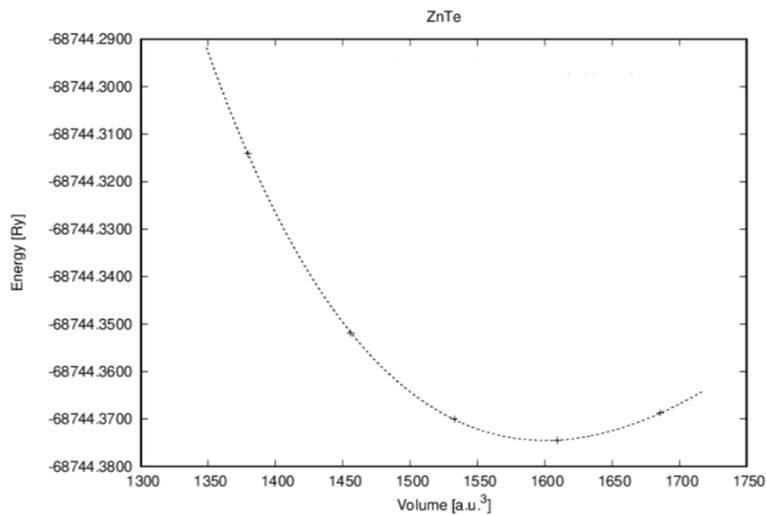

**Fig. 2**  Optimization of the lattice constant "a" of ZnTe compound

## 3.2  Electronic properties

To determine the electronic band structures for ZnTe, the conduction band (CB) and the valence band (VB) calculated in directions of high symmetry of the first Brillouin zone are given in **Fig. 3** when applying the correction of Tran and Blaha modified Becke-Johnson (TB-mBJ) approach using the Wien2K code.

We know that the energy gap is defined as the difference between the maximum of the valence band and the minimum of the conduction band with the maximum of the valence band and the minimum of the conduction band located at the symmetry point Γ. Its mean direct gap at the high symmetry points Γ-Γ whose value is 2.16eV. According to the band structure, we have a gap energy Eg = 2.16eV and we can see that the valence band is closer to the Fermi energy which implies that we have a p-type semiconductor.

To study the total and partial density of ZnTe states, we have plotted the obtained results in **Fig. 4** (a) and (b). The density of states of the material (**Fig. 4** (a) and (b)) shows that the contribution of the p orbital of the Te anion is dominant near the valence band maximum with a very small contribution from the two s and p orbitals of the Zn cation. From the curve, we can see that the up spins are symmetrical with the down spins which implies that this material is non-magnetic.

Then, ZnTe is a semiconductor (no overlap between the valence and conduction band) because it has a gap energy of about 1eV.

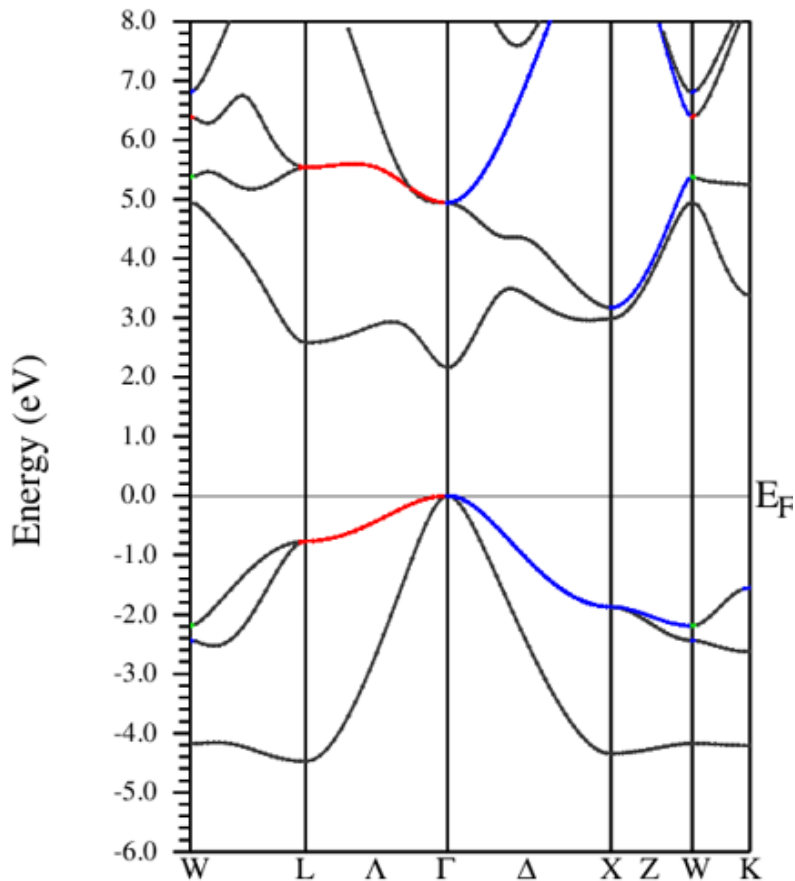

**Fig. 3** Band structures of ZnTe.

.

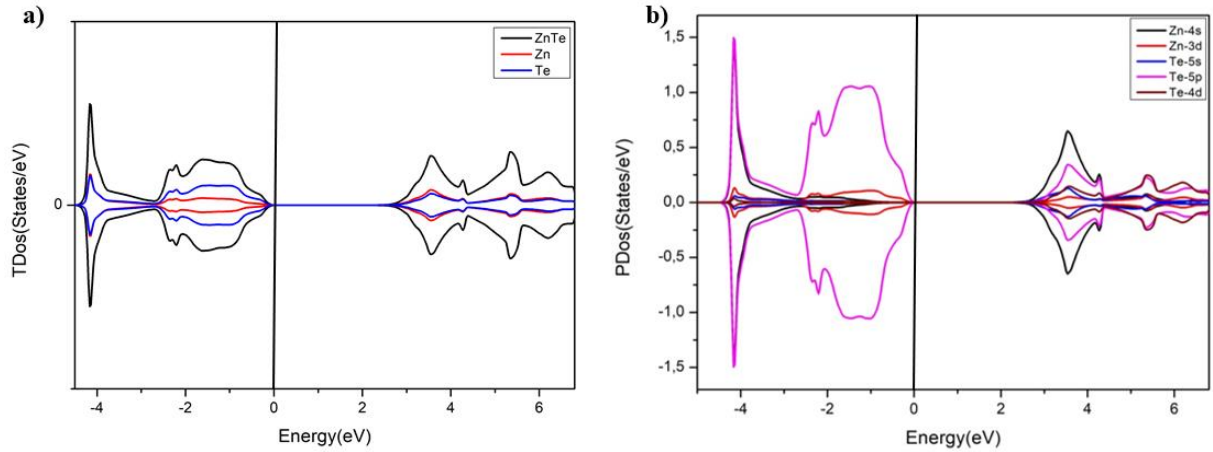

**Fig. 4** Total and partial density of states of ZnTe. In (a) the total DOS. In (b) the partial DOS

.

## 3.3 Optical properties for ZnTe

In this section, we will present the optical properties of ZnTe namely absorption, reflectivity, and refraction using ab initio methods implemented in the Wien2k code using the GGA approximation.

The optical responses of the ZnTe phase are calculated for photon energies up to 13eV. **Fig. 5** shows the real and imaginary parts of the dielectric (absorption) function of the ZnTe compound.

From the spectrum of the real part $\varepsilon_1(\omega)$ one can see that the real part changes slowly with the increase of the energy to reach in the ultraviolet region a maximum value equal to 7.2eV, while the positive values of the real part represent the propagation of light inside the material.

The spectrum of the imaginary part $\varepsilon_2(\omega)$ shows a first peak located at 4.8 eV, followed by three distinguished located at 6.5 eV, 9.5 eV, and 11.8 eV, respectively. Using the band structure and the density of state started with the paragraph above, we have identified these different peaks. The first peak generally comes from the direct transition of valence electrons from the Zn-4s, 3d, and Te-5s, 5p states and to the Zn-4s, 3d, and Te-5s, 5p states of the lower part of the conduction band along the Γ- Γ direction.

The second and third peaks are produced primarily from the direct (Γ-Γ) and indirect (Γ- L; Γ- X) transition of valence electrons to the Zn-(s, d) and Te-(s, p) orbits in the upper part of the conduction band. The last peak is probably due to electronic transitions from the Te-4d orbits located in the lower part of the valence band (-11 eV to -13 eV). It is interesting to note that the imaginary part remains zero below the energy 2.1 eV (the infrared-visible range). This reflects the absence of interactions between the medium and the incident photons.

**Fig. 6** shows the reflectivity spectra of ZnTe at zero frequencies, the reflectivity is about 18%. Then, it increases to reach maximum values in the Ultra-Violet range.

We can see from **Fig. 7** that the refractive index has an increasing trend of low energy. It changes very little in the infrared-visible region, then grows rapidly in the ultraviolet region until it reaches a maximum value equal to 3.51 at about 3.5 eV. This high value of n (ω) tells us that ZnTe behaves as an opaque material for this incident electromagnetic wave. At high energy (more than 25 eV) the refractive index stabilizes towards 0.

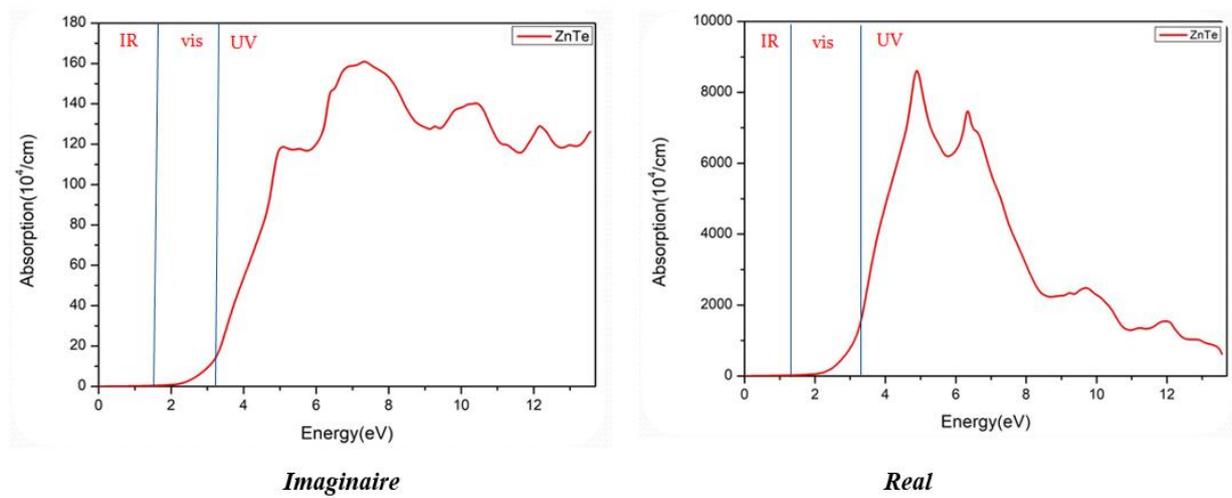

*Imaginaire*            *Real*

**Fig. 5** Absorption spectrum of ZnTe as a function of photon energy eV.

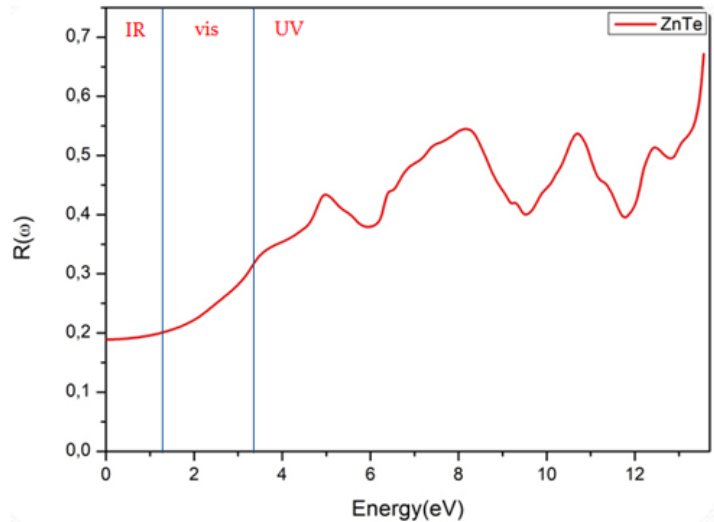

**Fig. 6** reflectivity curve of ZnTe as a function of photon energy eV.

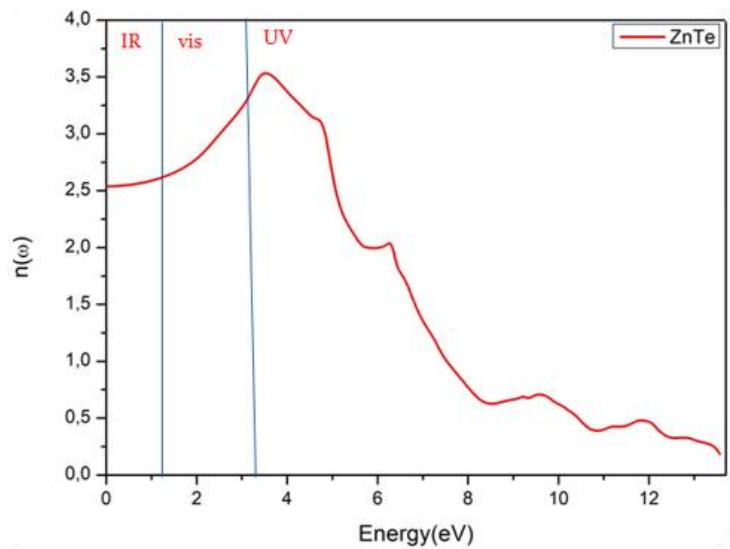

**Fig. 7** Refraction curve of ZnTe as a function of photon energy eV.

### 3.4 Doping of ZnTe with Mn

In this part, we will take the material ZnTe by doping it with a transition metal, we use here the element Mn to have the magnetization in this material.

**Fig. 8** shows the structure of ZnTe doped by Mn (ZnMnTe), so we can see that we have a structure Zinc-Blende, with the Mn atom occupying the vertexes of the cube and the Zn atom occupying the centers of faces and the Te atoms occupying the tetra sites of the cube.

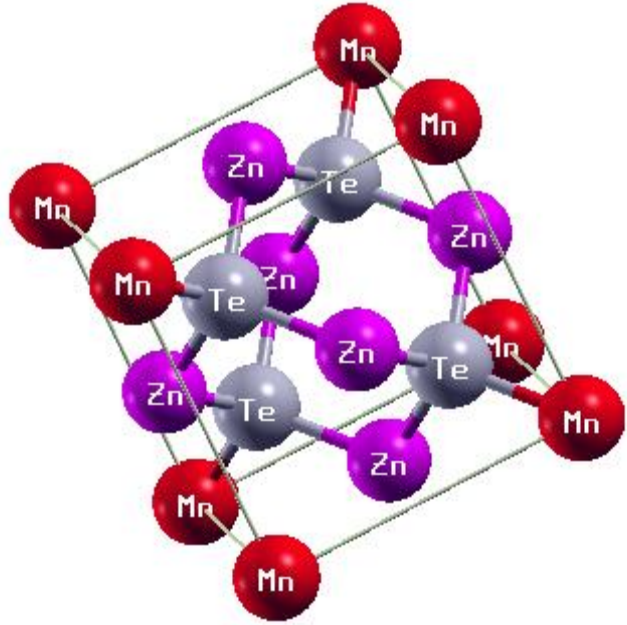

**Fig. 8** Structure of ZnTe doped with Mn

.

## 3.5 Electronic properties of ZnMnTe

**Fig.9** shows the energy band structures of the ZnMnTe compounds. The band structure of the up spins has a metallic behavior because the valence band crosses (touches) the Fermi energy. And the band structure of the down spins has a semiconductor behavior of type n because we have a more important value of the energy of gap around Eg=1.72 eV, and its nature is a direct gap. Also, we have the minimum of the conduction band closer to the Fermi level which confirms the n-type of this semiconductor.

From **Fig. 10** in the energy region of the valence and conduction band, the 3d orbitals of the Mn atom are denser than the orbits of the other atoms because this orbital is responsible for the magnetism of the material considered. As for **Fig. 11**, we can observe that we have a magnetic behavior because we don't have symmetry between the up spins and the down spins. And we see that the gap between the valence and conduction bands is equal to Eg=1.72eV.

We compare these results with the one obtained before doping and we notice that the gap is decreased because of doping with the magnetic atom Mn which explains the metallic behavior that we found.

So, after these results and the results obtained in the band structure, we can say that the compound ZnMnTe is semi-metallic.

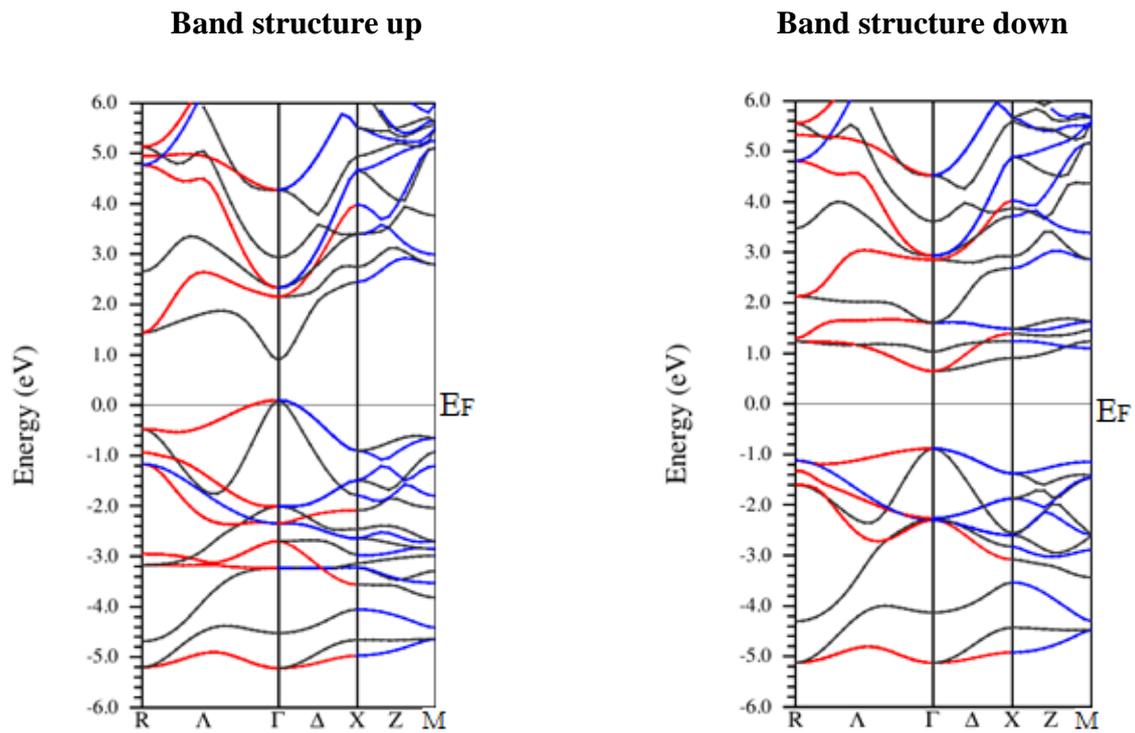

**Fig. 9** Band structures of ZnMnTe for Spin-up and Spin-down

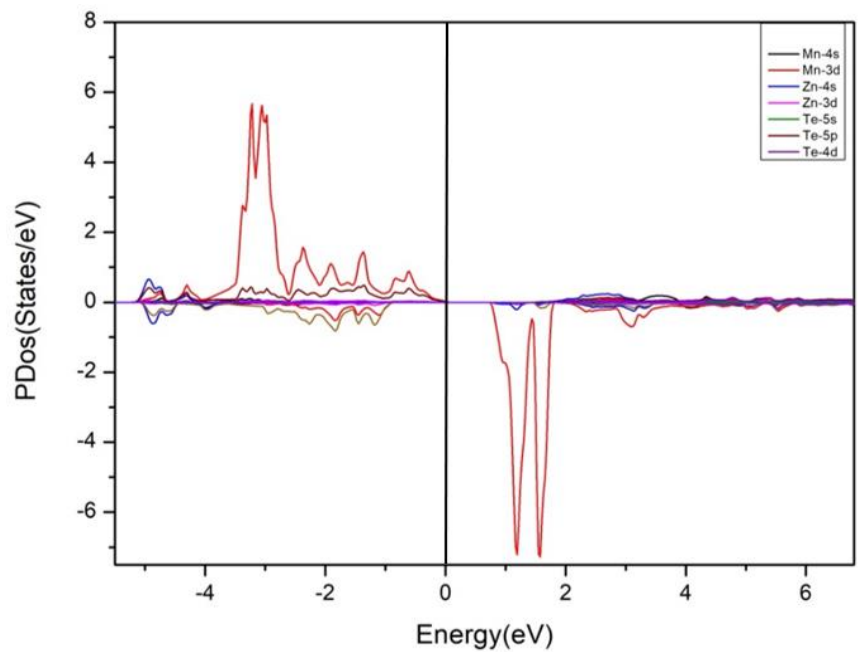

**Fig. 10** Partial density of states of ZnMnTe

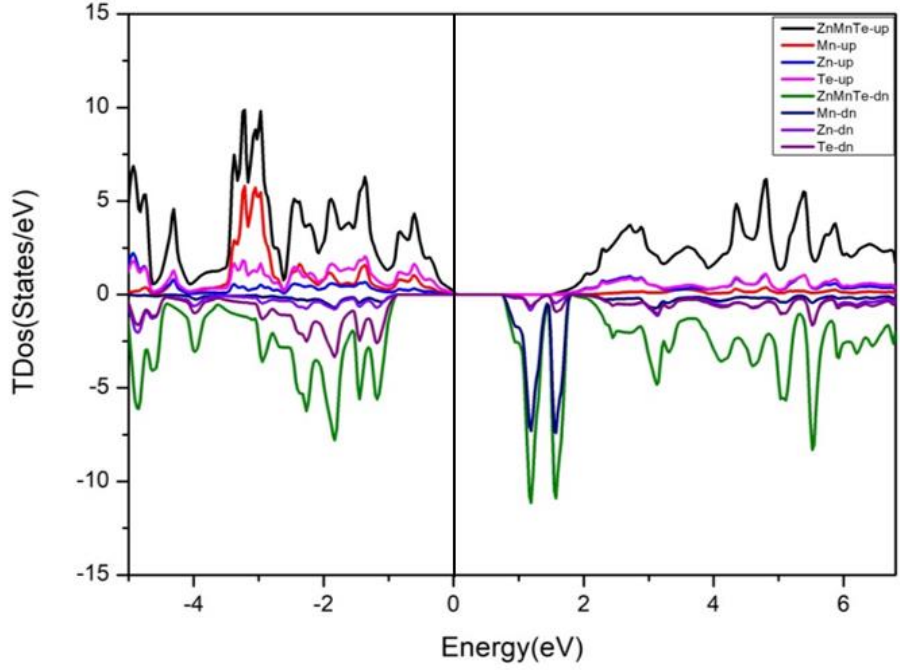

**Fig. 11** Total density of states of ZnMnTe

## 4 Monte Carlo simulations

In this part, we applied Monte Carlo simulations (MCS) to study the magnetic properties of the ZnMnTe compound for non-zero temperature values. Such simulations were performed with the Metropolis algorithm [35-37]. The Hamiltonian governing the system under study is given by:

$$H = -J \sum_{<ij>}^{z} S_i S_j - \Delta \sum_{i=1}^{N} S_i S_i - h \sum_{i}^{k} S_i \qquad (1)$$

where the notation <ij> represents the spin of the first nearest neighbor. J is the magnetic exchange coupling interaction between the Mn-Mn atoms. The parameters $\Delta$ and h are the crystalline and external magnetic fields, respectively. The spin moment of $Mn^{2+}$ ions is $S = \frac{3}{2}$. This means that the $S_i$ variables in Eq. (1) take the values $\{+\frac{3}{2}; +\frac{1}{2}; -\frac{1}{2}; -\frac{3}{2}\}$.

The J-exchange coupling interaction of the ZnTe compound is studied by first-principal calculations at T = 0 K using Eq. (2):

$$J = \frac{E(antif) - E(ferro)}{S^2 Z} \qquad (2)$$

where E (ferro), and E (anti) are the energies of the ferromagnetic and antiferromagnetic phases, respectively. Z = 6 is the number of nearest neighbor ions, see **Fig. 8** The value obtained is:

J=-2.18 meV, see **Table 1**.

The crystal field is expressed as:

$$\Delta = \frac{E_a}{\sum (S_i)^2} \tag{3}$$

where the magnetic anisotropic energy $E_a$ is equal to the energy difference between t2g and eg the energy positions of the orbitals.

The magnetic properties of compound ZnMnTe are studied using Monte Carlo calculations based on the Metropolis algorithm. Cyclic constraints imposed on the lattice were forced and configurations were generated by successively traversing the lattice and performing single-spin flip experiments. Also, we choose a random spin in the system and consider its inverse. Then it is important to calculate the energy variation between the two states from the Hamilton expression (Eq. (1)). If the energy difference is negative, the spin-flip is accepted. Otherwise, if the energy variation is positive, the spin changes if the randomly chosen number are less than the Boltzmann probability: $e^{-\beta H}$.

The following results present the variations of magnetization (M), susceptibility ($\chi$), specific heat ($C_v$) as well Binder Cumulant ($U_L$). For the different sizes 4, 8, 16, and 32. At spin round configuration, we perform $2\times10^4$ Monte Carlo steps and discard the first $5\times10^3$ configurations generated. First, for each iteration, we start by calculating the internal energy per site given by Eq. (4). The energy, magnetization, magnetic susceptibility, heat energy, and the Binder Cumulant have been respectively computed by using the following equations:

$$E = \frac{1}{N} H \tag{4}$$

$$M = \frac{1}{N} \sum_{i}^{N} |S_i| \tag{5}$$

$$\chi = \frac{1}{K_B * T} (\langle M^2 \rangle - \langle M \rangle^2) \tag{6}$$

$$C_v = \frac{1}{K_B * T^2} (\langle E^2 \rangle - \langle E \rangle^2) \tag{7}$$

$$U_L = 1 - \frac{\langle M^4 \rangle_l}{3 \langle M^2 \rangle_l^2} \tag{8}$$

where $k_B$ is the Boltzmann constant. The total number of spins is $N = L \times L \times L$. T is the absolute temperature.

To study the magnetic properties of the ZnMnTe compound, we performed Monte Carlo simulations using the Metropolis approach. We record the behavior of magnetization, susceptibility, and specific heat as a function of temperature, for a fixed value of the magnetic exchange coupling (J = -2.18meV) and the crystalline field is negligible because it is very small ($\Delta$ = 0 meV), with no external magnetic field (h = 0) in Figs. 12-13-14-15. The three images obtained show that the material ZnMnTe undergoes a phase transition at the Neel Temperature $T_N$ = 154.31 **K**.

The results of the variation of magnetization with temperature for the different sizes of the system (L = 4, 8, 16, and 32) are shown in **Fig. 12**. We notice a phase change, and when the temperature increases, the magnetization decreases until it disappears at a critical temperature $T_c$ = 6.1 J/$K_B$ (Neel Temperature).

We have a first-order transition because the variation of the magnetization shows a discontinuity at Tc.

The measurement of the susceptibility is based respectively on the fluctuations of the magnetization of the system. Furthermore, **Fig. 13** represents the variation of the susceptibility as a function of temperature. The susceptibility diverges in the vicinity of a temperature that indicates the phase transition; its value tends towards the exact value when we tend towards the thermodynamic limits.

The specific heat is a quantity that makes it possible to measure the temperature according to the energy, and its variation according to the temperature is represented in **Fig. 14** for various sizes; We can see that the peaks correspond to the critical temperature are increased with the size, as well as, the specific heat is absent until the temperature reaches the 3 values, then it starts to diverge near the critical temperature, and after that the decrease tends to 0 value with increasing temperature.

A more numerically efficient method is the determination of the Binder Cumulant "cumulant of order four". From **Fig. 15** we can confirm that: we change the size of the system all the points meet at the same point which is the point of transition (Critical Temperature → Neel Temperature).

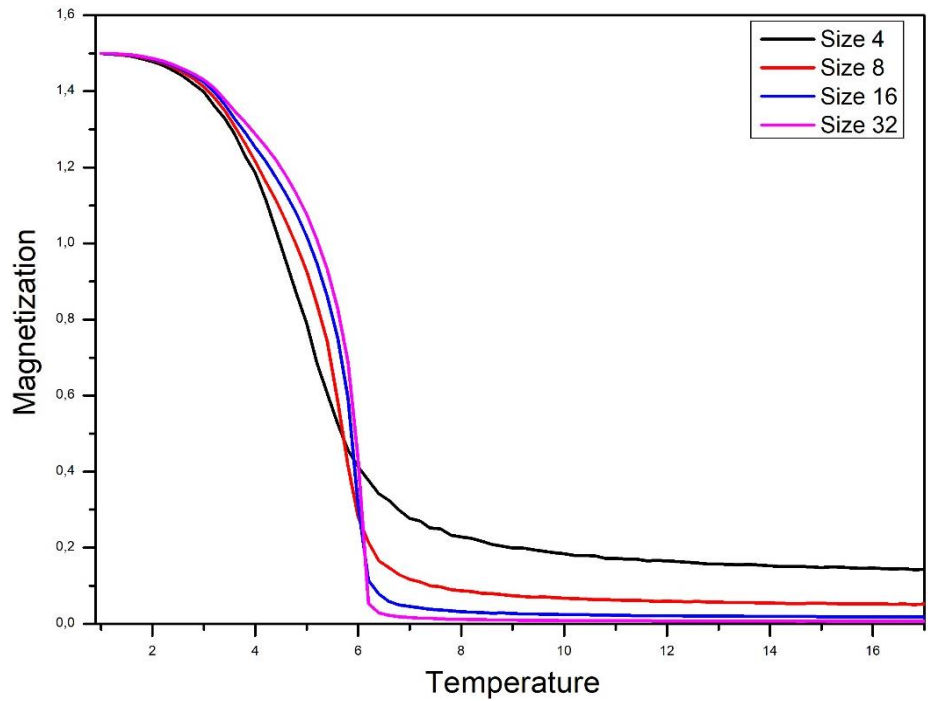

**Fig. 12** The thermal variation of the magnetization for different sizes of the material ZnMnTe

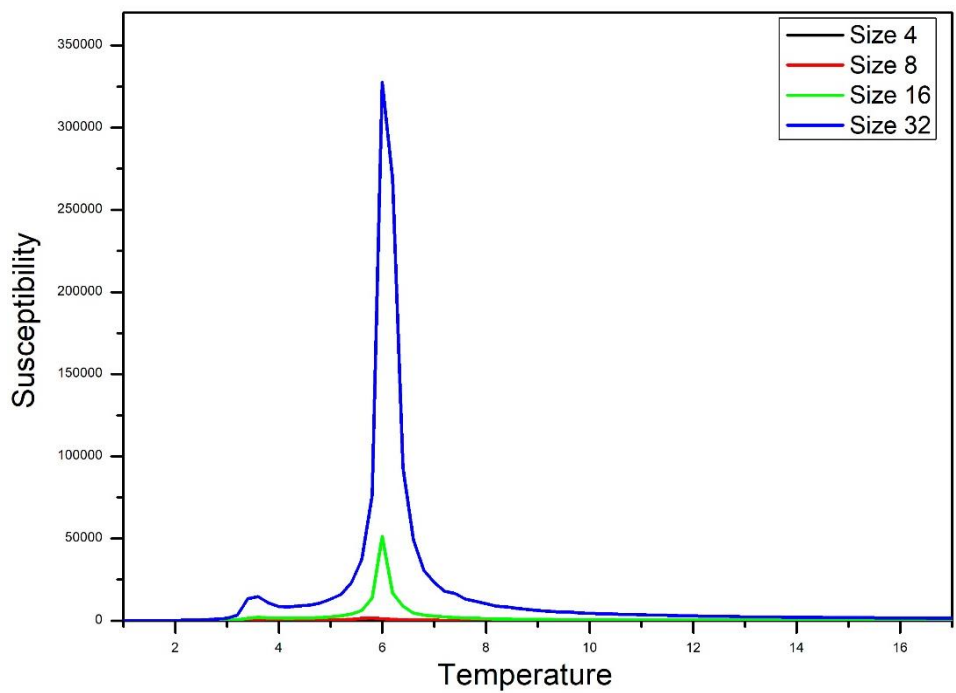

**Fig. 13** The thermal variation of susceptibility for different sizes for the material ZnMnTe

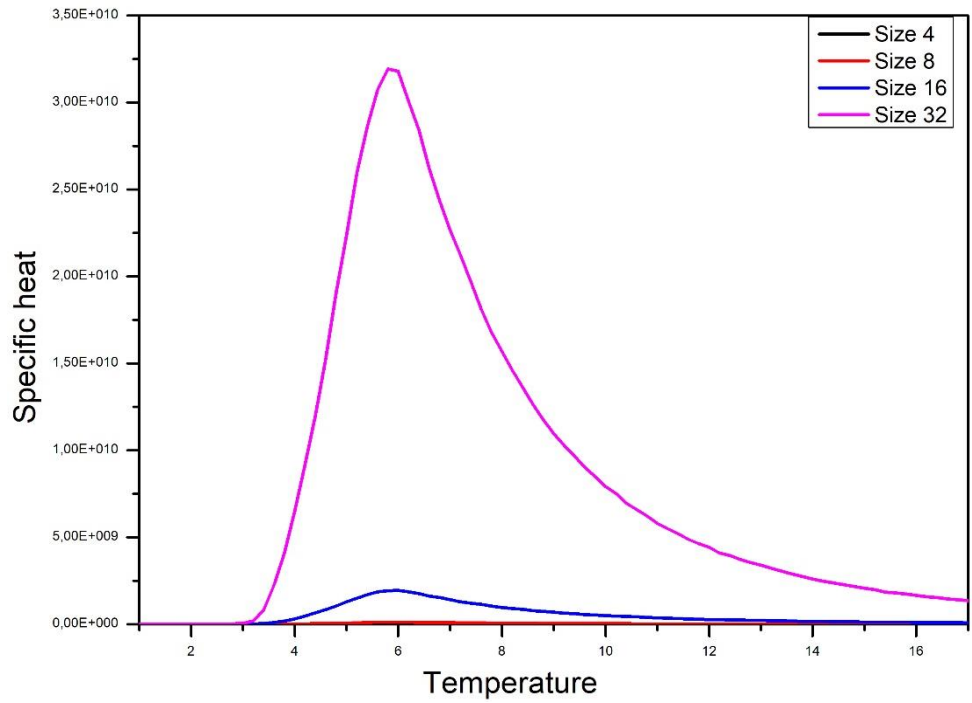

**Fig. 14** The thermal variation of specific heat for different sizes for the material ZnMnTe

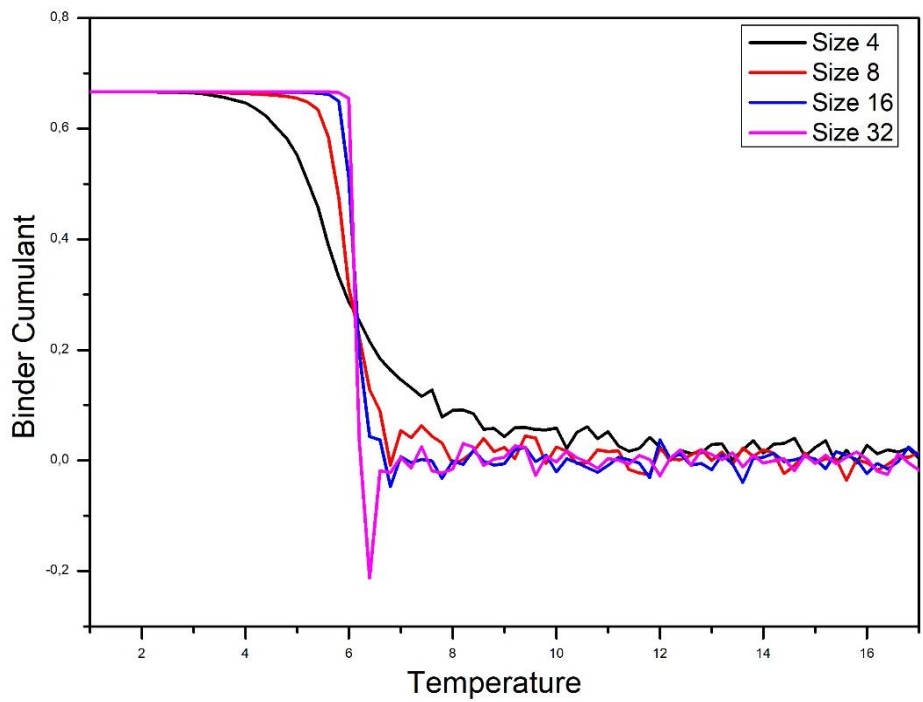

**Fig. 15** The thermal variation of Binder Cumulant for the material ZnMnTe

**Table 1** Calculated ferromagnetic, antiferromagnetic, total energies, and coupling exchange interaction by ab initio calculation

| E(ferro) (eV) | E(antif) (eV) | E (eV) | J (meV) |
|---|---|---|---|
| -1835843,054 | -1835843,113 | -0,0589 | -2,18 |

# 5 Conclusion

In this article, we applied ab initio calculations using the GGA approximation to simulate the structural, electronic, optical, and the magnetic properties of the compound ZnTe. Through the Use of the GGA approximation, we notice that the ZnTe band structure has semiconductor character since there is no band overlap between the valence and conduction band, and there is also a noticeable gap Eg=2.16eV between them. Then, we took the material ZnTe and doped it with a transition metal Mn to have the magnetization in that material and we observe that the latter has gained the semimetal character. However, we use Monte Carlo simulations to study the magnetic properties of the studied compound (magnetization, specific heat, and susceptibility). The findings attained by Monte Carlo simulations display that the ZnMnTe material goes through an antiferromagnetic phase to the paramagnetic phase at the Neel Temperature $T_N$ =159.31 K.

## References

[1] Uwatoko, Y., Umehara, I., Ohashi, M., Nakano, T., & Oomi, G. (2012). Thermal and electronic properties of rare earth compounds at high pressure. In Handbook on the Physics and Chemistry of Rare Earths (Vol. 42, pp. 1-164). Elsevier.

[2] Steinfeld, A., & Meier, A. (2004). Solar fuels and materials. In Encyclopedia of energy (Vol. 5, pp. 623-637). Elsevier Academic Press.

[3] Zhao, Y., He, D., Qian, J., Pantea, C., Lokshin, K. A., Zhang, J., & Daemen, L. L. (2005). Development of high P–T neutron diffraction at LANSCE–toroidal anvil press, TAP-98,

in the HIPPO diffractometer. In Advances in high-pressure technology for geophysical applications (pp. 461-474). Elsevier.

[4] Mark Yeadon, J.Murray Gibson, Molecular Beam Epitaxy, Semiconductors, Encyclopedia of Physical Science and Technology (Third Edition), Academic Press, 2003, Pages 113-121, ISBN 9780122274107, https://doi.org/10.1016/B0-12-227410-5/00457-9

[5] Hooker, C. (2011). Conceptualising reduction, emergence and self-organisation in complex dynamical systems. In Philosophy of complex systems (pp. 195-222). North-Holland.

[6] A.N. Semenov, I.A. Nyrkova, 1.02 - Statistical Description of Chain Molecules. Polymer Science: A Comprehensive Reference Volume 1, 2012, Pages 3-29

[7] BRÄUER, Gerhard, SACHSENHOFER, Klaus, et LANG, Reinhold W. Material and process engineering aspects to improve the quality of the bonding layer in a laser-assisted fused filament fabrication process. Additive Manufacturing, 2021, vol. 46, p. 102105.

[8] TANG, Ben Zhong, GENG, Yanhou, LAM, Jacky Wing Yip, et al. Processible nanostructured materials with electrical conductivity and magnetic susceptibility: preparation and properties of maghemite/polyaniline nanocomposite films. Chemistry of materials, 1999, vol. 11, no 6, p. 1581-1589.

[9] GOOVAERTS, Kristel, LAMBRECHTS, Paul, DE MUNCK, Jan, et al. Encyclopedia of materials: science and technology. Elsevier Science, 2002.

[10] . Jena, P., and A. W. Castleman Jr. "Introduction to atomic clusters." Science and Technology of Atomic, Molecular, Condensed Matter & Biological Systems. Vol. 1. Elsevier, 2010. 1-36.

[11] JUN, Sukky et LIU, Wing Kam. Moving least-square basis for band-structure calculations of natural and artificial crystals. In : Material Substructures in Complex Bodies. Elsevier Science Ltd, 2007. p. 163-205.

[12] K. Binder,Computer Simulation Techniques in Condensed Matter Physics, Encyclopedia of Condensed Matter Physics, Elsevier, 2005, Pages 211-219, ISBN 9780123694010, https://doi.org/10.1016/B0-12-369401-9/00564-7.

[13] KURTH, S., MARQUES, M. A. L., et GROSS, E. K. U. Density-Functional Theory, Editor (s): Franco Bassani, Gerald L. Liedl, Peter Wyder, Encyclopedia of Condensed Matter Physics. 2005.

[14] S.M. Blinder,Chapter 14 - Density functional theory,Introduction to Quantum Mechanics (Second Edition),Academic Press,2021,Pages 235-244.

[15] Jean-Paul Crocombette, François Willaime,1.16 - Ab Initio Electronic Structure Calculations for Nuclear Materials,Comprehensive Nuclear Materials (Second Edition),Elsevier,2020,Pages 517-543.

[16] VOGL, P., HJALMARSON, Harold P., et DOW, John D. A semi-empirical tight-binding theory of the electronic structure of semiconductors. Journal of physics and chemistry of solids, 1983, vol. 44, no 5, p. 365-378.

[17] J.-P. Crocombette, F. Willaime,1.08 - Ab Initio Electronic Structure Calculations for Nuclear Materials,Comprehensive Nuclear Materials,Elsevier,2012,Pages 223-248,

[18] CHAKRABORTY, Brahmananda. Electronic structure and theoretical aspects on sensing application of 2D materials. In : Fundamentals and Sensing Applications of 2D Materials. Woodhead Publishing, 2019. p. 145-203.

[19] HE, Zhengran, ZHANG, Ziyang, ASARE-YEBOAH, Kyeiwaa, et al. Crystal growth of small-molecule organic semiconductors with nucleation additive. Current Applied Physics, 2021, vol. 21, p. 107-115.

[20] HYUN, Jerome K. et ZHANG, Shixiong. Growth of nanowire heterostructures and their optoelectronic and spintronic applications. In : Magnetic Nano-and Microwires. Woodhead Publishing, 2020. p. 103-133.

[21] ALVIAL-PALAVICINO, Carla et KONRAD, Kornelia. The rise of graphene expectations: Anticipatory practices in emergent nanotechnologies. Futures, 2019, vol. 109, p. 192-202.

[22] KIM, Munho, SEO, Jung-Hun, SINGISETTI, Uttam, et al. Recent advances in free-standing single crystalline wide band-gap semiconductors and their applications: GaN, SiC, ZnO, β-Ga 2 O 3, and diamond. Journal of Materials Chemistry C, 2017, vol. 5, no 33, p. 8338-8354.

[23] MITZI, David B. Polymorphic one-dimensional (N2H4) 2ZnTe: Soluble precursors for the formation of hexagonal or cubic zinc telluride. Inorganic chemistry, 2005, vol. 44, no 20, p. 7078-7086.

[24] Azouzi, W., Sigle, W., Labrim, H., Benaissa, M., Sol-gel synthesis of nanoporous LaFeO3 powders for solar applications, Materials Science in Semiconductor Processing, 2019, 104, 104682.

[25] Ziti, A., Hartiti, B., Labrim, H. et al, Effect of copper concentration on physical, properties of CZTS thin films deposited by dip-coating technique, Applied Physics A: Materials Science and Processing, 2019, 125(3), 218.

[26] S. Idrissi, H. Labrim, L. Bahmad, A. Benyoussef, DFT and TDDFT studies of the new inorganic perovskite CsPbI3 for solar cell applications, Chemical Physics Letters, Volume 766, 2021, 138347, https://doi.org/10.1016/j.cplett.2021.138347.

[27] ZAARI, Halima. Etude ab initio des propriétés optiques des matériaux: cas du ZnTe, CdFe2O4, MgB2. 2015.

[28] Idrissi, S., Labrim, H., Ziti, S., Bahmad, L., A DFT study of the equiatomic quaternary Heusler alloys ZnCdXMn (X=Pd, Ni or Pt), Solid State Communications, 2021, 331, 114292

[29] Goumrhar, F., Bahmad, L., Mounkachi, O., Benyoussef, A., Ab-initio calculations for the electronic and magnetic properties of Cr doped ZnTe, Computational Condensed Matter, 2018, 15, pp. 15–20

[30] Goumrhar, F., Bahmad, L., Mounkachi, O., Benyoussef, A., Ab initio calculations of the magnetic properties of TM (Ti, V)-doped zinc-blende ZnO, International Journal of Modern Physics B, 2018, 32(3), 1850025

[31] BLAHA, Peter, SCHWARZ, Karlheinz, MADSEN, Georg KH, et al. wien2k. An augmented plane wave+ local orbitals program for calculating crystal properties, 2001, vol. 60, p. 1-302.

[32] ANDERSEN, O. Krogh. Linear methods in band theory. Physical Review B, 1975, vol. 12, no 8, p. 3060.

[33] MOMMA, Koichi et IZUMI, Fujio. VESTA 3 for three-dimensional visualization of crystal, volumetric and morphology data. Journal of applied crystallography, 2011, vol. 44, no 6, p. 1272-1276.

[34] Y.Yu, J. Zhou, H. Han, C. Zhang, T. Cai, C. Song, et T. Gao, Journal of Alloys and Compounds, 471,492, mars 2009.

[35] R. Masrour, E.K. Hlil, M. Hamedoun, A. Benyoussef, O. Mounkachi, L. Bahmad, Theoretical investigation of electronic and magnetic properties of MnAu layers, J. Magn. Magn Mater. 326 (2013) 166–170.

[36] MASROUR, R., JABAR, A., LABIDI, S., et al. Electronic, magnetic, reentrant and spin compensation phenomena in Fe2MnGa Heusler alloy. Physica Scripta, 2020, vol. 95, no 6, p. 065803.

[37] S. Kadri, S. Labidi, R. Masrour, A. Jabar, M. Labidi &M. Ellouze, Investigation of total and partial magnetic moments of Mn2NiAl with pressure at several temperatures, Phase Trans. J. 92 (8) (2019) 699–706.